\documentclass{article}
\usepackage{PRIMEarxiv}
\usepackage{authblk}
\usepackage[utf8]{inputenc} 
\usepackage[T1]{fontenc}    
\usepackage{hyperref}       
\usepackage{url}            
\usepackage{booktabs}       
\usepackage{amsfonts}       
\usepackage{nicefrac}       
\usepackage{microtype}      
\usepackage{lipsum}
\usepackage{fancyhdr}       
\usepackage{graphicx}       
\graphicspath{{media/}}     

\usepackage{amssymb}
\usepackage{tabularx}
\usepackage{threeparttable}
\usepackage{csquotes}
\usepackage{textgreek}
\usepackage{float}
\usepackage{siunitx}
\usepackage{stmaryrd}

\usepackage{physics}
\usepackage{multirow}

\usepackage{caption}
\usepackage{subcaption}
\usepackage{xcolor}

\usepackage{booktabs}
\usepackage{array} 
\usepackage{colortbl}
\usepackage{diagbox}
\usepackage{stmaryrd}
\usepackage{empheq, etoolbox}
\title{High-fidelity modeling of interface crossing in the diffusion welding process at the polycrystalline scale }

\author[1,2,3]{C. Godinot}
\affil[1]{CEA, LCA laboratory, Grenoble, France}
\author[1]{E. Rigal}
\author[3]{F. Bernard}
\author[1]{P. Emonot}
\author[1]{P.-E. Frayssines}
\author[1,2]{L. Védie}
\author[2]{M. Bernacki}
\affil[2]{Mines Paris, PSL University, Centre for material forming (CEMEF), UMR CNRS, 06904 Sophia Antipolis, France}
\affil[3]{Europe Burgundy University, ICB, UMR CNRS Dijon,France}
\begin{document}
\maketitle

\begin{abstract}
Controlling the microstructure of a diffusion welded interface is a critical point to ensure optimum mechanical properties and the homogeneity of the joint. Beyond the intimate contact formation between bonded parts studied in the literature, this article focuses on the grain boundary crossing of the interface during this process and its measurement. Following this perspective, a Level-Set method has been used for full-field microstructure simulations in 2D with various interface parameters. Two crossing measurement models have been formulated, tested and discussed over the simulations.
\end{abstract}


\section{Introduction}\label{introduction}
Hot isostatic pressing (HIP) is widely used in industry for sintering \cite{FunkenbuschHIPBiModalPowder1992}, densifying cast metals \cite{OnaHIPPingEffectsSteel1992}, and diffusion welding (also known as diffusion bonding) \cite{AtkinsonFundamentalaspectshot2000b} - a welding process in which the temperature remains below the melting point of the base materials. During this process, the parts to be welded are placed in a canister, which is then vacuumed before closure. The canister is subsequently placed in the HIP vessel, which is also closed to initiate the gas pressurization and heating. Typical HIP diffusion welding cycles last several hours, with a temperature level fixed between 70\% and 80\% of material melting point and a pressure range from several tens MPa to few hundred MPa.

One of the primary advantages of the HIP diffusion welding process is the ability to join several metallic parts intimately with a limited impact on the microstructure. Unlike other solid-state welding processes such as explosion welding or cold welding, HIP involves isostatic stress on the parts resulting in reduced material deformation. Furthermore, in an ideal scenario, the inital interfaces become indistinguishable from the core material due to interface healing and grain boundaries migration. 

However, in order to maintain a high yield stress in the material e.g. for fatigue resistance, the grain size must be carefully monitored according to the Hall-Petch effect \cite{LiHallPetcheffect2016}, \cite{CorderoSixdecadesHall2016}. While grain boundary migration is needed for the disappearance of the interface, this mechanism will also drive a potentially unwanted excessive grain growth. This last assertion is of course specific to the case where maximizing the yield stress is the main goal. Indeed, it will not be relevant for optimizing the creep strength where a medium to high grain size is preferable \cite{SherbyInfluencegrainsize2002}, for an opposite example.

\begin{figure}
	\centering 
	\includegraphics[width=1\textwidth]{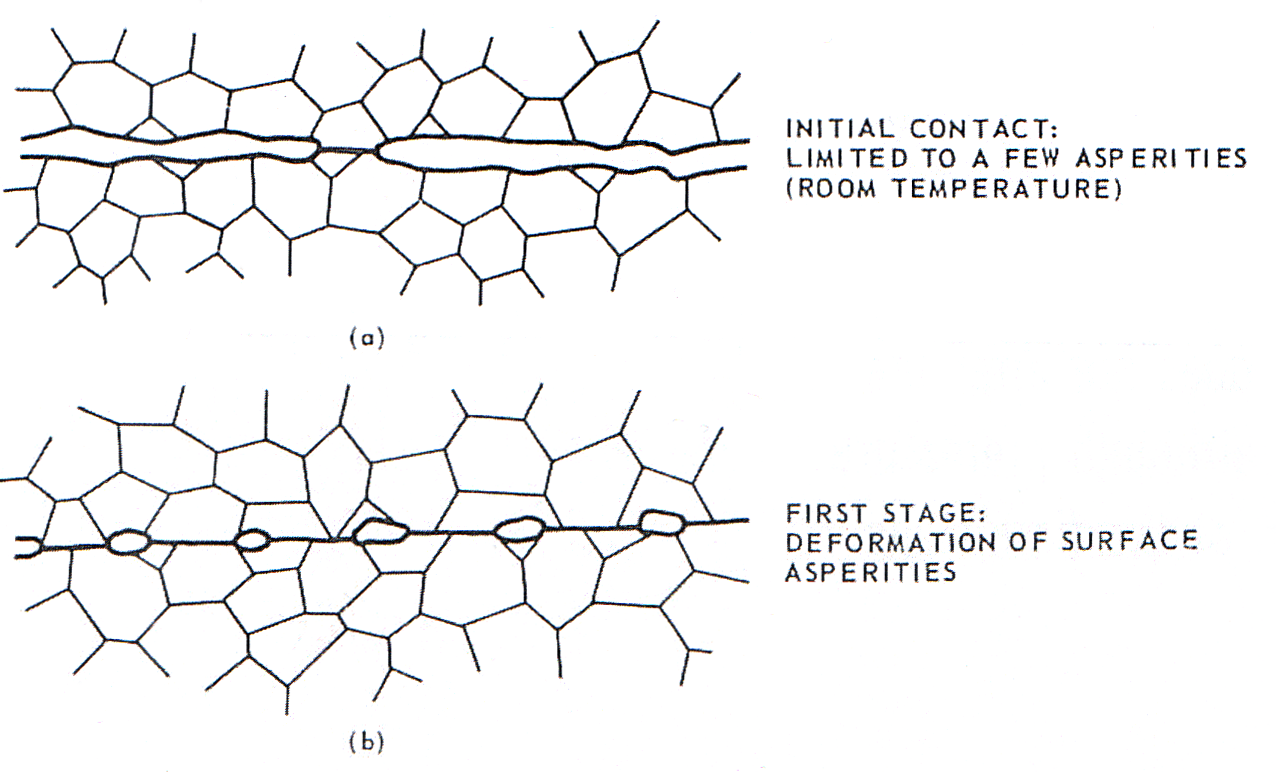}
    \includegraphics[width=1\textwidth]{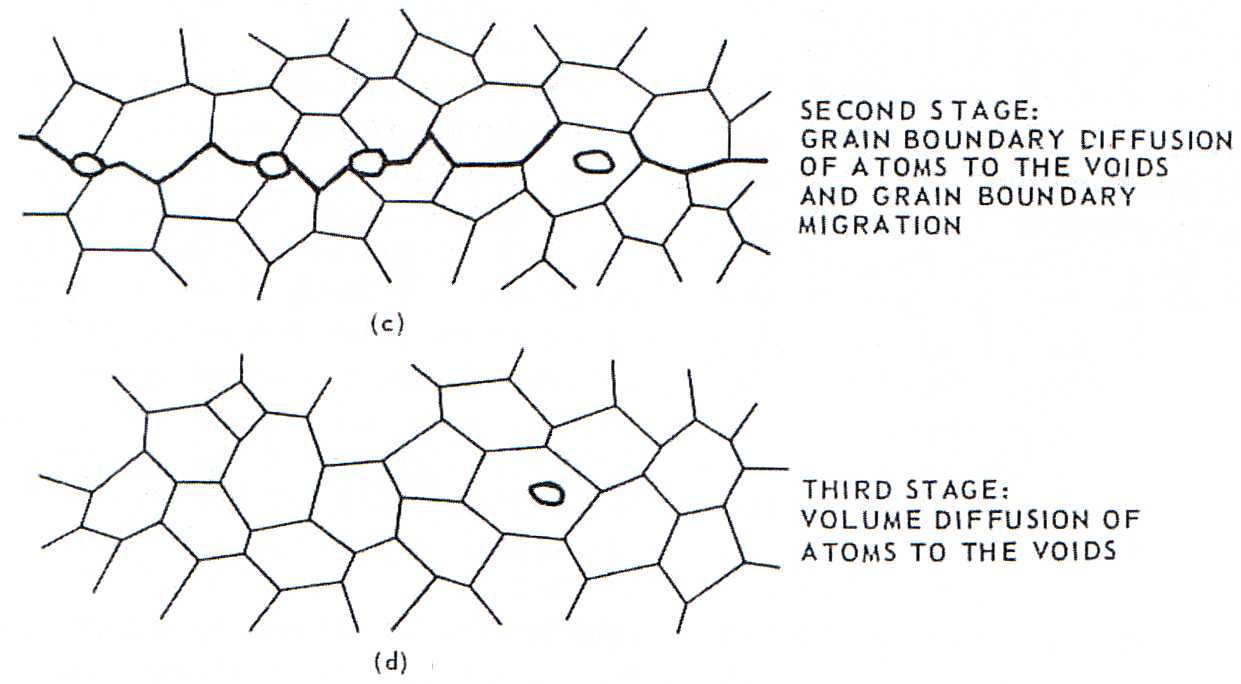}
	\caption{a) Initial state before bonding, b) Interface formation after plastic deformation, c) Interface crossing and voids shrinkage, and d) residual voids closure - Illustrative pictures from \cite{KingAdditionalStudiesDiffusion1968}.} 
	\label{fig:king1}
\end{figure}

As illustrated in Fig.\ref{fig:king1}, and based on experiments using Hot Uniaxial Pressing (HUP) for titanium alloys, the different stages of diffusion welding at the interface were clearly identified and described in 1954 by King and Owczarski \cite{KingAdditionalStudiesDiffusion1968}. While this process shares several features with HIP, it must be emphasized that in HUP, pressure is applied via a hydraulic press, leading to higher normal stresses on the parts and thus potentially greater strain. Nevertheless, the welding mechanisms remain the same.

The first stages (see Fig.\ref{fig:king1}) consists in the plastic and viscoplastic deformation as the stress (either isostatic or uniaxial) is progressively imposed on the parts with the temperature rise. The initial contact points between the parts (see Fig.\ref{fig:king1}(a)), which depend on the surface roughness, rapidly flatten, resulting in an interface with residual porosity (see Fig.\ref{fig:king1}(b)). The hardening is limited, the metal being rapidly restored or recrystallized at the beginning of the process. As the canister has been vacuumed before HIP, the interface voids shall not be confused with trapped gas.

The following stages includes both grain boundary migration by grain growth and pore closure by diffusion mechanisms (giving its name to the process). The newly formed grain boundaries move out of the interface plane, initiating crossing. At this stage, some grain boundaries may become unpinned from voids, while others moving toward voids can become pinned by these obstacles, consistent with the Smith-Zener pinning mechanism \cite{Zener1949,Smith1948}, which applies to all-types of suprananometric second phase particles. In the meantime, the interface voids shrink and spheroidize, mostly by both surface and volume diffusion mechanisms at the grain boundaries (see Fig.\ref{fig:king1}(c)). At this point, it is worth noting that while the (visco-)plastic deformation of the first stage is over once the applied stress drops below a fraction of the yield stress, grain boundary migration (and grain growth) will continue regardless of the crossing state. 

At the last stage, the voids intersecting a grain boundary have been closed and the remaining ones are located inside the grains. They are closed at a slower rate by volume diffusion, due to the lack of preferential diffusion paths such as grain boundaries (see Fig.\ref{fig:king1}(d)). 

Although this sequential description is helpful for understanding the main mechanisms involved in diffusion welding, it is important to note that the stages often overlap. For instance, grain boundary migration through recrystallization and grain growth can begin before the initial plastic deformation is complete, and both grain boundary and volume diffusion may occur simultaneously. This step-by-step description primarily highlights the order in which the dominant mechanisms tend to emerge.

Moreover precipitates can form at the interface during thermal cycle, creating new obstacles for grain boundary migration.

One of the first quantitative model dedicated to diffusion bonding was proposed by Hamilton \cite{HamiltonPressureRequirementsDiffusion1973} who correlated the imposed temperature/stress parameters of a HUP cycle to the fraction of welded area over time for titanium. While this first analytical approach was solely based on a viscoelastic deformation model, it allowed a first quantification of the time needed to obtain a given welding ratio at the interface. Garmong et al. used the same approach \cite{GarmongAttainmentfullinterfacial1975} to develop a slice model where the local stress is calculated for each slice. These initial qualitative models were followed by numerous studies on diffusion welding, which have been progressively refined over time.
Coble's work on sintering mechanisms \cite{CobleDiffusionModelsHot1970} has then served as a core foundation for the subsequent models of void closure \cite{GuoModellingdiffusionbonding1987, ChenDiffusivegrowthgrainboundary1981a,TakahashiRecentvoidshrinkage1992,Orhannewmodeldiffusion1999a}. 

The Derby and Wallach model \cite{DerbyTheoreticalmodeldiffusion1982} as well as the Hill and Wallach HUP model \cite{HillModellingsolidstatediffusion1989} consider seven distinct mechanisms in competition for void closure during the whole diffusion bonding process. In these approaches, a 2D cylindrical void is considered, on which each mechanism contribution is formulated depending on the initial interface geometry (no difference is made between grain boundaries connected to the void and the welding interface) and the applied stress. The ellipse major semi-axis and neck radius (itself derived from major and minor semi-axis) are computed to follow the evolution of void closure. Analytical formulations and contributions of each mechanisms for void closure are discussed in a review of Y. Takahashi and K. Inoue \cite{TakahashiRecentvoidshrinkage1992} in order to delimit the predominance domain for each one. 

All these state of the art studies generally focus on the closure of the interface, but grain boundary migration is not taken into account. At best, a mean grain size parameter is fixed for the whole process in order to calculate the interface diffusion coefficients. The welding ratio only partly describes the welding state of the parts, as the final material strength is not quite the same whether the crossing is complete or not \cite{BouquetEtudeformationjoints2014,Nurnberg316LStainlessSteel2017}. As the focus of this article is made on the interface crossing, the different models mentioned above will not be discussed more thoroughly.

Most precisely, high-fidelity modeling of grain growth enabling to consider a second phase particle population if of interest here. In this context, numerous full field modeling have been proposed, including front-tracking/vertex \cite{WeygandZenerpinninggrain1999, Couturier2003, Florez2020, Florez2022, Florez2025}, Monte Carlo/Cellular Automata methods \cite{Srolovitz1984,Hassold1990,Gao1997,Kad1997,Phaneesh2012,Villaret2020}, level-set (LS) \cite{Agnoli2012, Agnoli2014a, Scholtes2016b, Villaret2020, Alvarado2021a, Alvarado2021b, Bernacki2024} and multiphase fields (MPF) approaches \cite{Chang2009, Tonks2015, Moelans2006, Chang2014}.  Front-capturing approaches like LS and MPF methods can reproduce precisely the local interaction between pore/grain interfaces and GB. In the LS framework, the concept of incorporating inert second phase particle within a finite element (FE) framework was initially proposed for conducting grain growth (GG) simulations \cite{Agnoli2012, Scholtes2016b, Villaret2020}, static recrystallization simulations \cite{Agnoli2014a} and also extend to dynamic particle interfaces \cite{Alvarado2021a, Alvarado2021b}. This approach enables the consideration of pores without predefined assumptions about their size or morphology. This approach will then be considered in the following.

\section{Modeling interface crossing by grain boundaries during diffusion bonding}
\label{sec:1}
\subsection{Hypothesis and assumptions}
\label{ssec:1}

In the present article, the focus is made on the second stage of the King and Owczarsky \cite{KingAdditionalStudiesDiffusion1968} description for the numerical modeling and simulation. The initial state of our model is considered to be fully recrystallized after plastic deformation, with no residual plastic stored energy and a totally bonded interface. These hypotheses are coherent for the welding by HIP of clean surfaces with fine polishing where initial (visco-)plastic deformation step duration can be considered as negligible compared to the grain growth mechanism duration. 
In order to take into account the influence of surface defaults (surface pollution, second phase particles, remaining voids...), obstacles will later be introduced. 
The initial generated microstructure is composed of equiaxed grains following a Rayleigh distribution in diameter and the grain boundary energy is estimated as an homogeneous mean value over all the microstructure. The grain boundary migration velocity $\vec{v}$ is assumed to follow the curvature flow due to the minimization of the global surface energy. Thus, by assuming an homogeneous grain boundary energy $\gamma$, the usual local kinetics law depends on the local mean curvature $\kappa$, on the temperature dependant mobility $\mu$ and on the grain boundary energy $\gamma$ through Eq.\ref{equ:velocity}:

\begin{equation}
\label{equ:velocity}
\vec{v} = -\mu\gamma\kappa\vec{n},
\end{equation}
with $\vec{n}$ the outside unitary normal to the grain boundary.

The mobility is classically assumed as a function of the absolute temperature following an Arrhenius law of activation energy $Q$ and pre-exponential factor $\mu_0$ through Eq.\ref{equ:Arrhenius}:
\begin{equation}
\label{equ:Arrhenius}
\mu\left(T\right) = \mu_0 \exp{\left(\frac{-Q}{RT}\right)}.
\end{equation}

The representativeness of Eq.\ref{equ:velocity} is increasingly debated in recent studies \cite{Chen2020,Bhattacharya2021,Florez2022,qiu2025}. These works ultimately emphasize that grain growth is not just curvature flow and that this equation remains a first order approximation. Nevertheless, at the polycrystalline scale, it remains a statistically excellent approximation, as extensively documented in the literature through numerous comparisons between experimental data and high-fidelity mesoscopic simulations.

\subsection{Level-set framework for polycrystal generation and grain growth simulation}
\label{ssec:2}

\begin{figure}
	\centering 
	\includegraphics[width=0.6\textwidth]{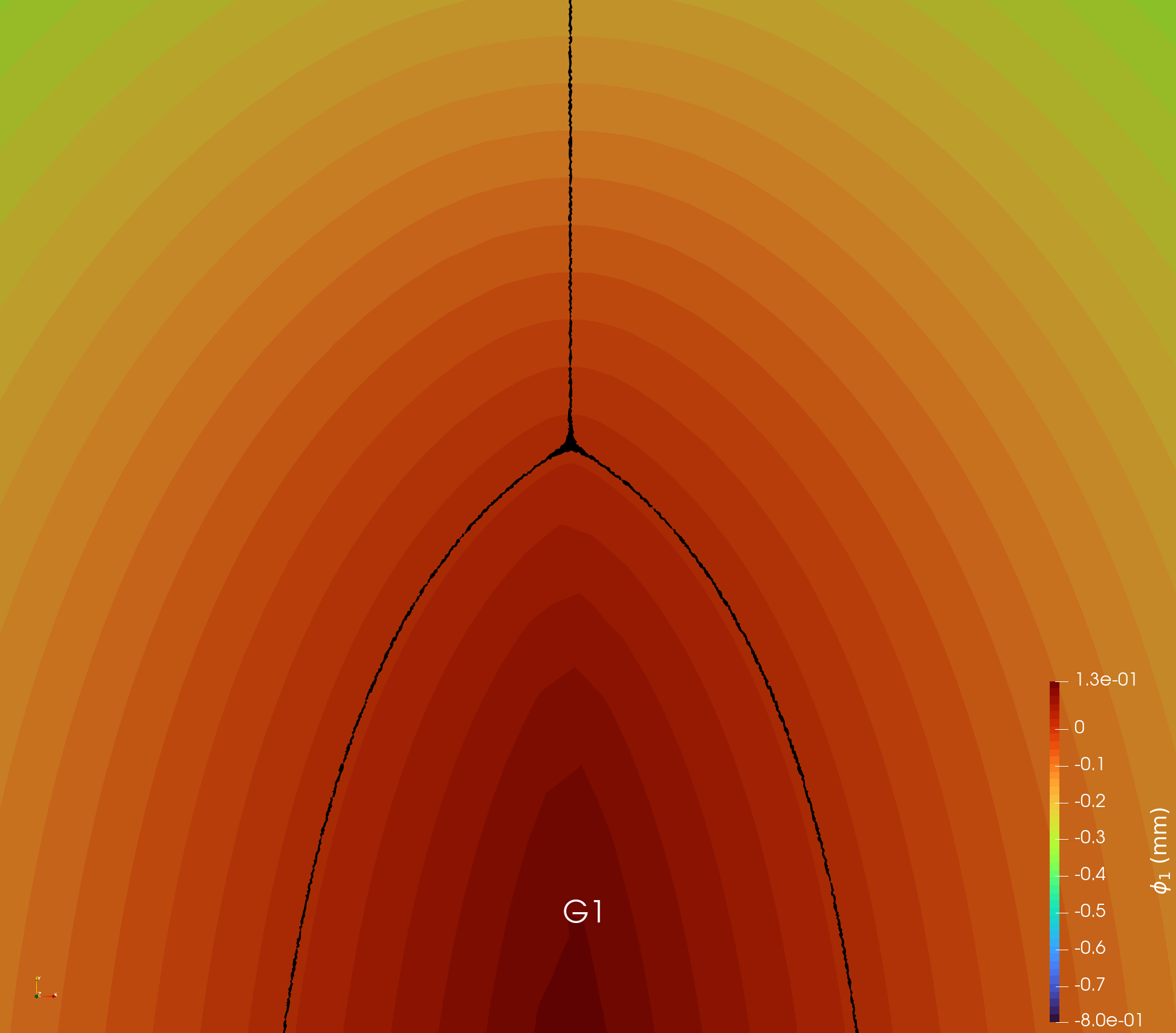}	
	\caption{Illustration of a triple junction in a LS context where the black lines correspond to the grain interfaces. The field corresponds to the distance function $\phi_1$ of the $G_1$ grain.  }
\label{fig:conventionLS}
\end{figure}

The simulations described hereafter are performed using a finite-element level-Set framework dedicated to the grain growth modeling in presence of a second phase \cite{Agnoli2012,Alvarado2021a,Bernacki2024}. The level-set function $\phi_i$ to a given grain $i \in \{1,2,...,N_{G}\}$ - positive inside the grain, negative outside by convention (see Fig.\ref{fig:conventionLS}) - defined over the Representative Volume Element (RVE) $\Omega$ is used to track its boundaries over time, the zero isovalue marking the position of the interfaces. 

Local outside normal unitary vector to the grain boundaries (GB) $\overrightarrow{n}_i$ and the mean curvature $\kappa_i$ (trace of the curvature tensor) at the interface can then be defined with differential operators applied to function $\phi_i$:
\begin{equation}
\label{equ:differential}
\overrightarrow{n}_i = -\cfrac{\overrightarrow{\nabla} \phi_i}{||\overrightarrow{\nabla} \phi_i||},\ \kappa_i = \overrightarrow{\nabla}.\overrightarrow{n}_i .
\end{equation}
When the considered level-set functions remain Euclidian distance function, i.e. $||\overrightarrow{\nabla} \phi_i|| = 1$, thus $\kappa_i = -\Delta\Phi_i$ and the classical convection equation to describe grain boundary migration thanks to the level-set description and through Eq.\ref{equ:velocity}:
\begin{equation}\label{eq:Transport}
\left \{
\begin{aligned}
    \partial_t \phi_i\left(\mathbf{x},t\right) + \mathbf{v}_i\left(\mathbf{x},t\right) \cdot \nabla \phi_i\left(\mathbf{x},t\right)=0\\
    \phi_i\left(\mathbf{x},t=0\right)=\phi_{i}^{0}\left(\mathbf{x}\right)
\end{aligned}
\right.
\end{equation}
with $\phi_{i}^{0}\left(\mathbf{x}\right)$ the initial distance function to the boundaries of the $i^{th}$-grain, can be rewritten introducing then the differential formulation for $\kappa$:
\begin{equation}\label{equ:LS0}
\left \{
\begin{aligned}
    \partial_t \phi_i\left(\mathbf{x},t\right) -\mu\gamma\Delta\phi_i\left(\mathbf{x},t\right)=0\\
    \phi_i\left(\mathbf{x},t=0\right)=\phi_{i}^{0}\left(\mathbf{x}\right)
\end{aligned}
\right. \qquad \qquad \forall i \in \{1,2,...,N_{G}\}.
\end{equation}

One level-set function can be used to track several non connected grains, reducing the calculus load with the number of level-set functions \cite{Elsey2009,Scholtes2015}. Special attention should be paid to the reinitialization/redistancing of the level-set functions after numerical resolution, as they must remain Euclidean distance functions almost near the grain boundaries \cite{Shakoor2015,Shakoor2025}. Finally, the curvature formulation presented won't be effective on multiple junctions, requiring a particular numerical treatments where different strategies coexist \cite{Zhao1996,Merriman1994,Li2025}. The review \cite{Bernacki2024} may interest readers curious about a comprehensive description of these different topics.

\begin{figure}[ht!]
  \centering
  \begin{subfigure}{0.9\textwidth}
    \centering
    \includegraphics[width=0.5\textwidth]{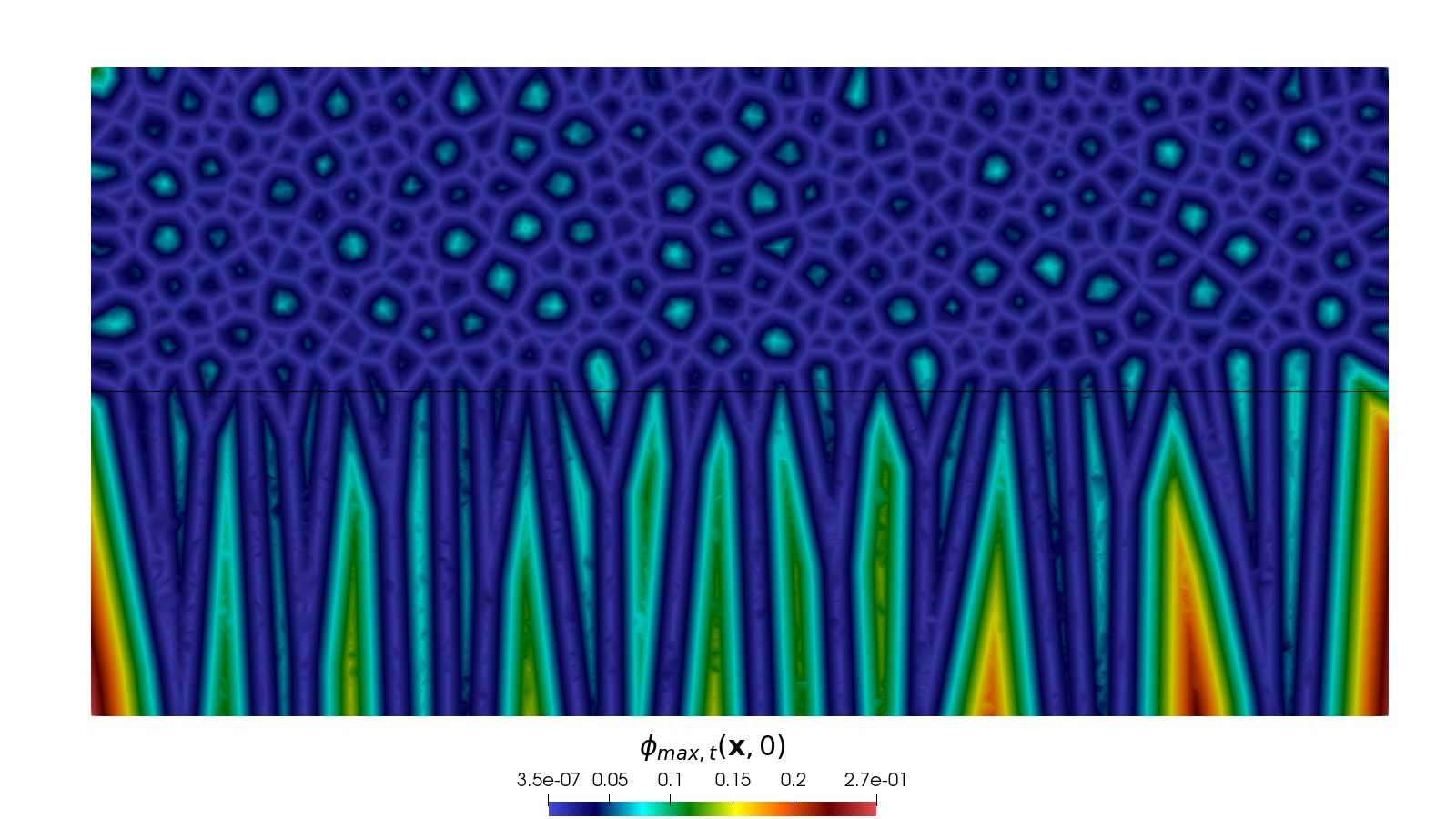}
    \caption{}
    \label{fig:Poly_t}
  \end{subfigure}
  \begin{subfigure}{0.9\textwidth}
    \centering
    \includegraphics[width=0.5\textwidth]{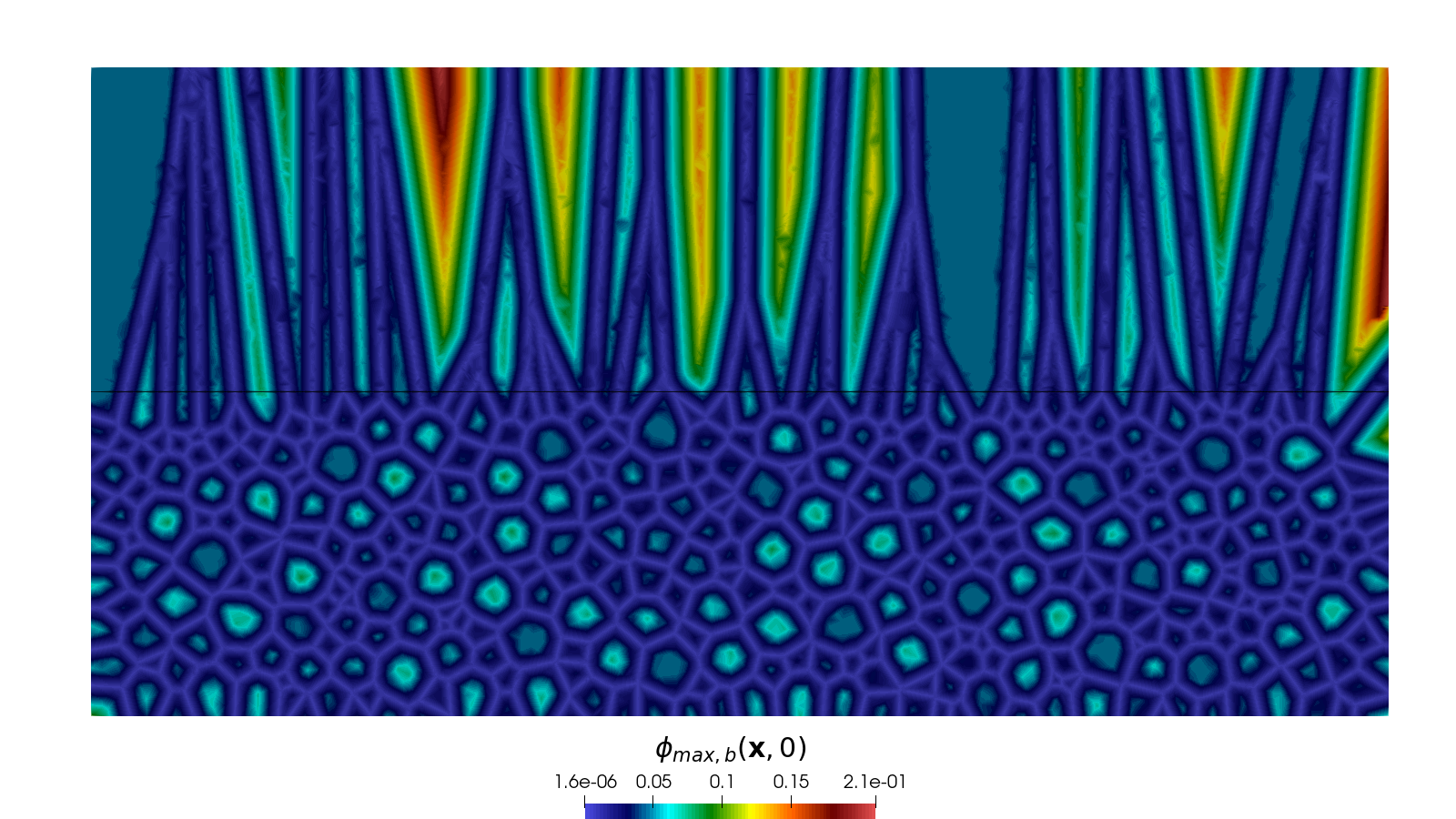}
    \caption{}
    \label{fig:Poly_b}
  \end{subfigure}
   \begin{subfigure}{0.9\textwidth}
    \centering
    \includegraphics[width=0.5\textwidth]{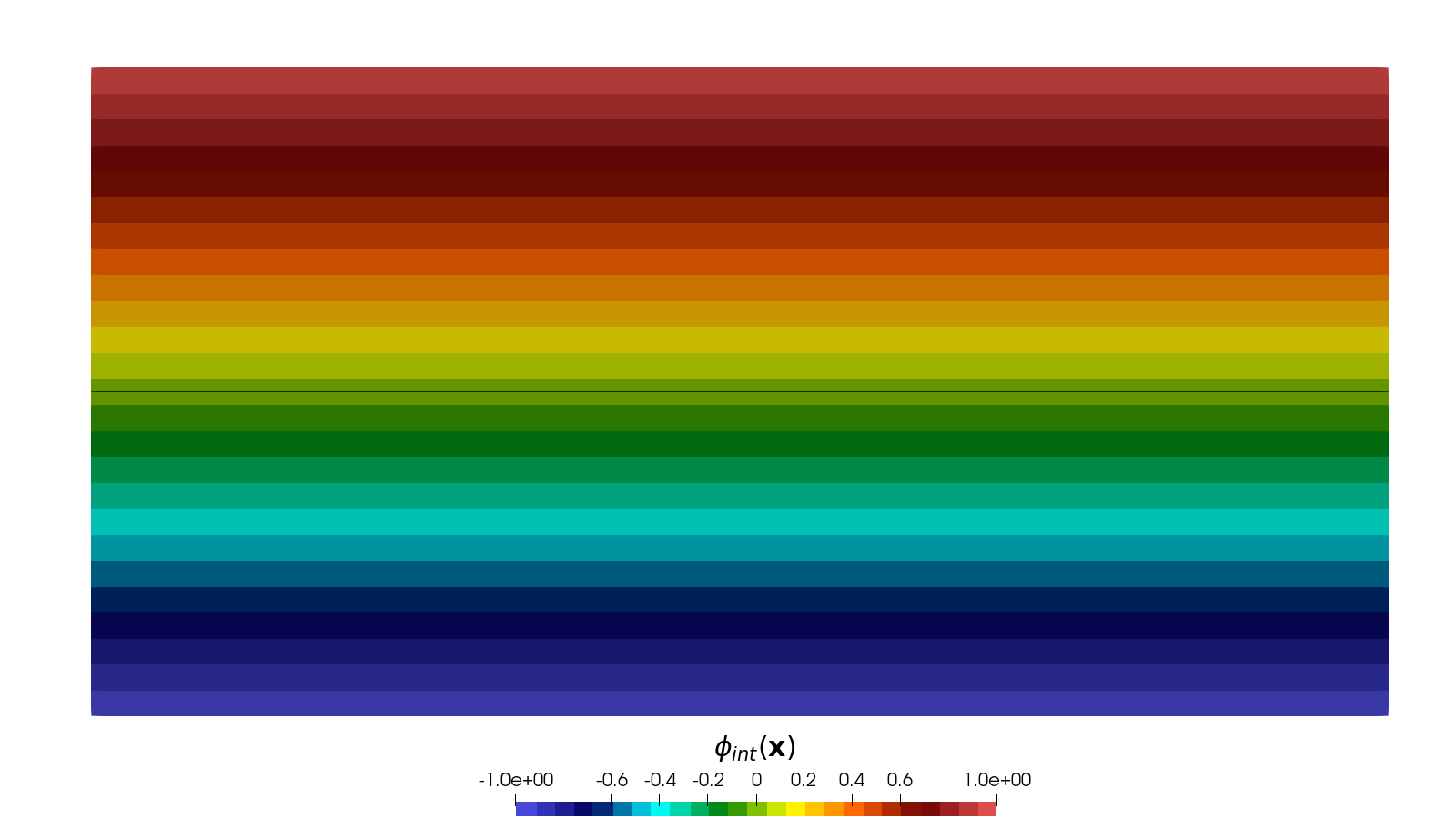}
    \caption{}
    \label{fig:Poly_i}
  \end{subfigure}
  \begin{subfigure}{0.9\textwidth}
    \centering
    \includegraphics[width=0.5\textwidth]{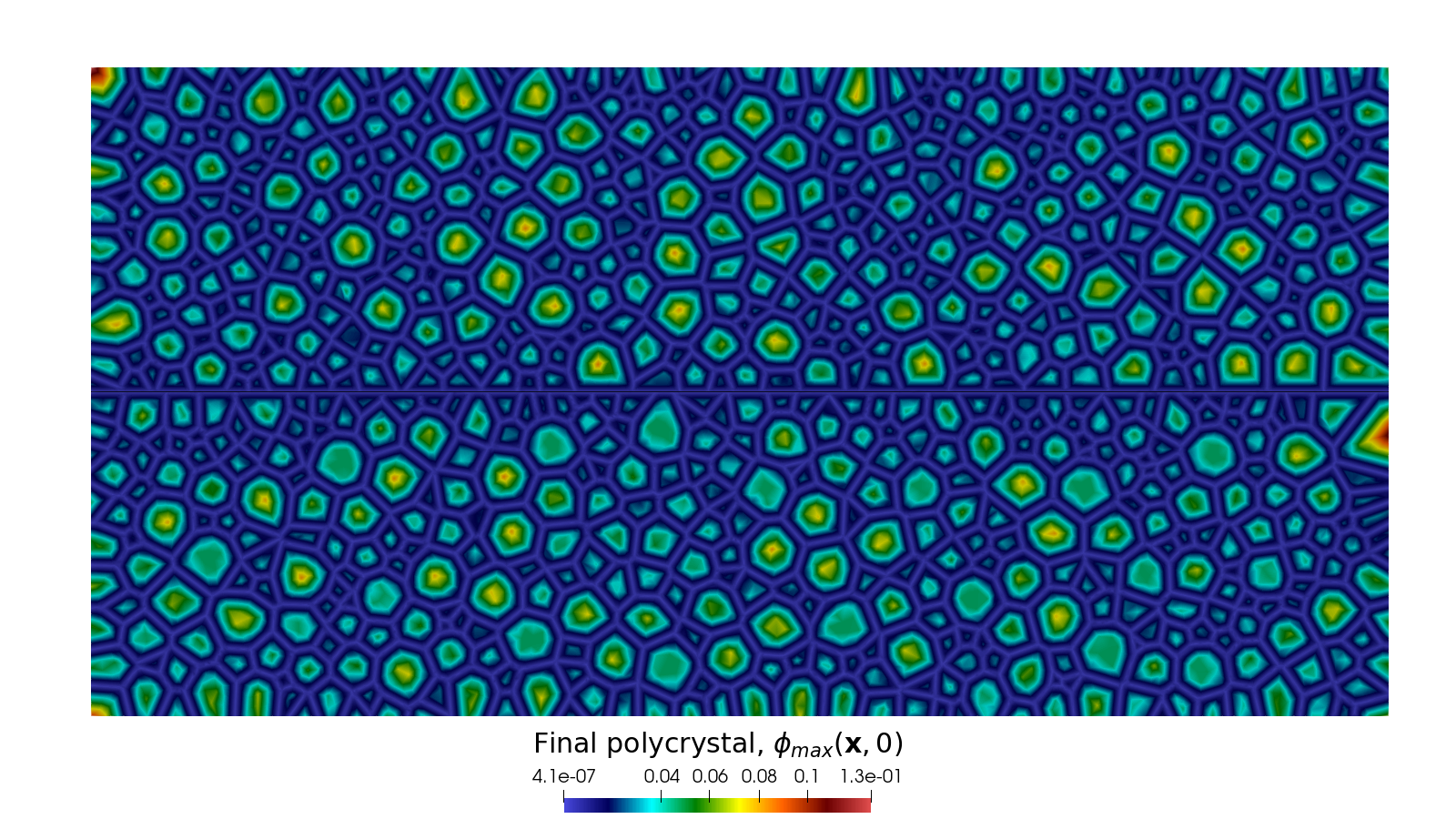}
    \caption{}
    \label{fig:Poly_p}
  \end{subfigure}

  \caption{Strategy for the generation of initial RVE for diffusion bonding simulation without obstacles: (a) polycrystal of the top part, (b) polycristal of the bottom part, (c) distance function to the bonding interface and (d) final polycrystal with the bonding interface.}
  \label{fig:DM}
\end{figure}

In the following, the initial polycrystals are generated by a drop-roll dense sphere packing method followed by a Laguerre-Voronoï generation \cite{Hitti2012,Hitti2013}. This method allows to control precisely the initial grain size distribution. In the present case, homogeneous diffusion welding simulation will require two generations, one for each domains separated by the bonding interface. The based-LS methodology is quite simple and illustrated in Fig.\ref{fig:DM} in a dimensionless $\left[0,4\right]\times \left[0,2\right]$ domain. Two polycrystals, following a Rayleigh law, are generated, respectively, in the top ($\left[0,4\right]\times \left[1,2\right]$) and bottom ($\left[0,4\right]\times \left[0,1\right]$) parts of the calculation domain. The figures \ref{fig:Poly_t} and \ref{fig:Poly_b} represent, respectively, the field $\phi_{max,t}\left(\mathbf{x},0\right)=\max_{i}\left(\phi_{i,t}\left(\mathbf{x},0\right)\right)$ and $\phi_{max,b}\left(\mathbf{x},0\right)=\max_{j}\left(\phi_{j,b}\left(\mathbf{x},0\right)\right)$ where $\phi_{i,t}$ and $\phi_{j,b}$ corresponds, respectively, to the level-set functions of the top, resp. bottom, polycrystal. By considering $\phi_{int}\left(\mathbf{x}\right)$ the distance function to the initial bonding interface as illustrated in Fig.\ref{fig:Poly_i} All the distances function can be modified as $\phi_{i,t}\left(\mathbf{x},0\right)= \min\left(\phi_{i,t}\left(\mathbf{x},0\right),\phi_{int}\left(\mathbf{x}\right)\right)$ $\phi_{i,b}\left(\mathbf{x},0\right)= \min\left(\phi_{i,b}\left(\mathbf{x},0\right),-\phi_{int}\left(\mathbf{x}\right)\right)$, leading by merging all the distance functions $\phi_{max}\left(\mathbf{x},0\right)= \max_{i,j}\left(\phi_{i,b}\left(\mathbf{x},0\right),\phi_{j,t}\left(\mathbf{x},0\right)\right)$ to the initial polycristal described in Fig.\ref{fig:Poly_p}. An identical result would be obtained by initially considering two polycrystals generated over the entire computational domain. An alternative to this global generation approach relies on the generation of substructures. Indeed, using the method proposed by Grand et al. \cite{Grand2022}, the bonding interface can be seen as the interface between two Laguerre–Voronoi cells (top and bottom parts), within which Laguerre–Voronoi subcells can be generated.

\subsection{Introduction of voids or incoherent second phase particles}
\label{ssec:3}

Several methods exist to model the interaction of second phase particles (SPP) with grain boundaries during grain growth. For static particles over time, one solution is to describe the second phase particles as holes in the domain. The particles are therefore not meshed and Neumann boundary conditions are applied at the interface. This simple method is detailed in \cite{Agnoli2014a,Bernacki2024}. However, this particle management has several limitations. Indeed, deformation simulations are difficult in the presence of second phase particles because they are not described and included in the mesh. In addition, particle evolutions cannot be taken into account. It is therefore not possible to simulate dissolution, growth, coalescence, spheroidization, or ripening of these SPPs. To be able to take these elements into account, Alvarado et al. proposed a new formalism in which the particles are represented by a level-set function  \cite{Alvarado2021a, Bernacki2024}. In the following, the first approach was used for simulation cases with static SPP and the second one for simulation cases with evolutive obstacles.\\

The obstacles, that can represent residual pores or second phase particles, are generated on a narrow zone around the bonding interface with a minimum distance between obstacles. An algorithm of the drop and roll type was favored as for the Laguerre-Voronoi algorithm descrived previously \cite{Hitti2013}. Depending on the discussed obstacles, their geometry, size, linear density and evolution differ. The choice of obstacle shape and size for this study will be detailed in section \ref{ssec:2.2}.

\section{Simulation processing and parameters selection}
\label{sec:2}

\subsection{Initial reduced mobility calibration}
\label{ssec:2.1}

The product of the mobility $\mu$ and interface energy $\gamma$ is referred to as reduced mobility : as the two parameters are not dissociated in Eq.\ref{equ:LS0} as in Eq.\ref{equ:velocity}, the kinetic calibration shall be performed over this single parameter. Here, the reduced mobility calibration is based on the grain size evolution datasets during heat treatment for a nuclear grade 316L steel.
Pure grain growth simulations (without interface) with analog initial grain size distribution and temperature over time have been done to iteratively adjust the $\mu\gamma$ parameter.

Following the hypothesis of Burke and Turnbull \cite{burke_recrystallization_1952}, the mean grain radius $R(t)$ is expected to evolve according to the following law referred as to the classical Burke-Turnbull equation, with $R_0$ the initial mean radius and $\alpha$ a constant :
\begin{equation}
\label{equ:BnT}
R^2(t) - R_0^2(t) = \alpha M\gamma t.
\end{equation}
In the literature, the modified Burke-Turnbull equation (see Eq.\ref{equ:BnT}) reformulates this equation as a more general law to overcome the limitation due to hypothesis linked to the initial model. It must be emphazised that the introduced exponent $n$ has no physical meaning regarding the initial Burke and Turnbull theory: 
\begin{equation}
\label{equ:BnTm}
R^2(t) - R_0^2(t) = \alpha M\gamma t^n.
\end{equation}

2D simulations indicates that the evolution of the ASTM grain size follows a classical Burke-Turnbull law (Eq.\ref{equ:BnT}), allowing the calibration to be made with a simple linear adjustment of the $\mu\gamma$ coefficient : for an isothermal heat treatment , considering a non-calibrated coefficient $(\mu\gamma)_0$, and an experimental dataset $D_\textrm{exp}$, $t_\textrm{exp}$ with $D_\textrm{exp}$ being the mean ASTM grain size measured at time $t_\textrm{exp}$, the calibrated reduced mobility $\mu\gamma$ is calculated from Eq.\ref{equ:calimobi}
\begin{equation}
\label{equ:calimobi}
(\mu\gamma)_1 = \cfrac{t_\textrm{exp}}{t_0},
\end{equation}
with $t_0$ the simulation time for which the value $D_\textrm{exp}$ is reached. In order to consider evenly each available ($D_\textrm{exp}$,$t_\textrm{exp}$) data in the calculus of an optimal reduced mobility coefficient, least square methods was applied for each temperature.



\subsection{Crossing criteria}

While it is acknowledged in the literature that the welding strength depends on the crossing of the grains at the interface in second order of priority (the first order being the voids closure), this crossing hasn't been really discussed in the state of the art before the works of Bouquet et al. \cite{BouquetEtudeformationjoints2014} where a normed criterion $C_\textrm{int}\left(\varepsilon\right)$ was tested over experimental datasets and was based on the measurement of the remaining grain boundary cumulative length $L_\textrm{int}$ within a thickness of $\varepsilon$ pixels around the initial interface compared to the total interface length $L_\textrm{tot}$:
\begin{equation}
\label{equ:cross0}
C_\textrm{int}\left(\varepsilon\right) = \cfrac{L_\textrm{int}(\varepsilon)}{L_\textrm{tot}}\\
\end{equation}

With this criterion, a zero value corresponds to an absence of crossing whereas a value close to one indicates a complete one. This criterion presents however some strong limits. First, it varies depending on the observation scale. Second, this criterion implies that the perfectly crossed state would be the total absence of grain boundaries inside a thin strip on the thickness $\varepsilon$. That means the crossing rate also depends on the grain size of the polycrystal. To eliminate the grain size dependency, the crossing state can be adjusted as $C_1\left(\varepsilon\right)$ (see Eq.\ref{equ:cross1}) where the previous crossing state calculated on the interface is divided by the one calculated on a random line in the RVE which does not cross the interface. This reference line has to be far enough from the limit of the domain to not be impacted by it. Last, if the interface has remaining pores or particles intersecting the bonding plane, their length have to be taken into account in the criterion. The interface portions containing precipitates are counted as crossed when the precipitates is in intragranular position and as not crossed otherwise.

\begin{equation}
\label{equ:cross1}
C_1\left(\varepsilon\right) = \frac{C_\textrm{int}\left(\varepsilon\right)}{C_\textrm{ref}\left(\varepsilon\right)}.\\
\end{equation}

In order to avoid the dependence to $\varepsilon$, a second criterion $C_2$ was here considered through Eq.\ref{equ:cross2} which compares the grain number at the interface plane $N_\textrm{int}$ with the one at a reference straight line $N_\textrm{int}$ located in the core microstructure. Initially, both surfaces put in close contact at in the interface are composed of grains traversed by the cut line. The lesser the difference in grain number between both lines, the closer the interface looks like the core microstructure. This criterion could highlight a grain size heterogeneity near the interface if it appears through the simulations. 

\begin{equation}
\label{equ:cross2}
C_2 =\frac{N_\textrm{ref}}{N_\textrm{int}}.
\end{equation}

\subsection{Results processing}

\begin{figure}
  \centering
    \includegraphics[width=0.8\textwidth]{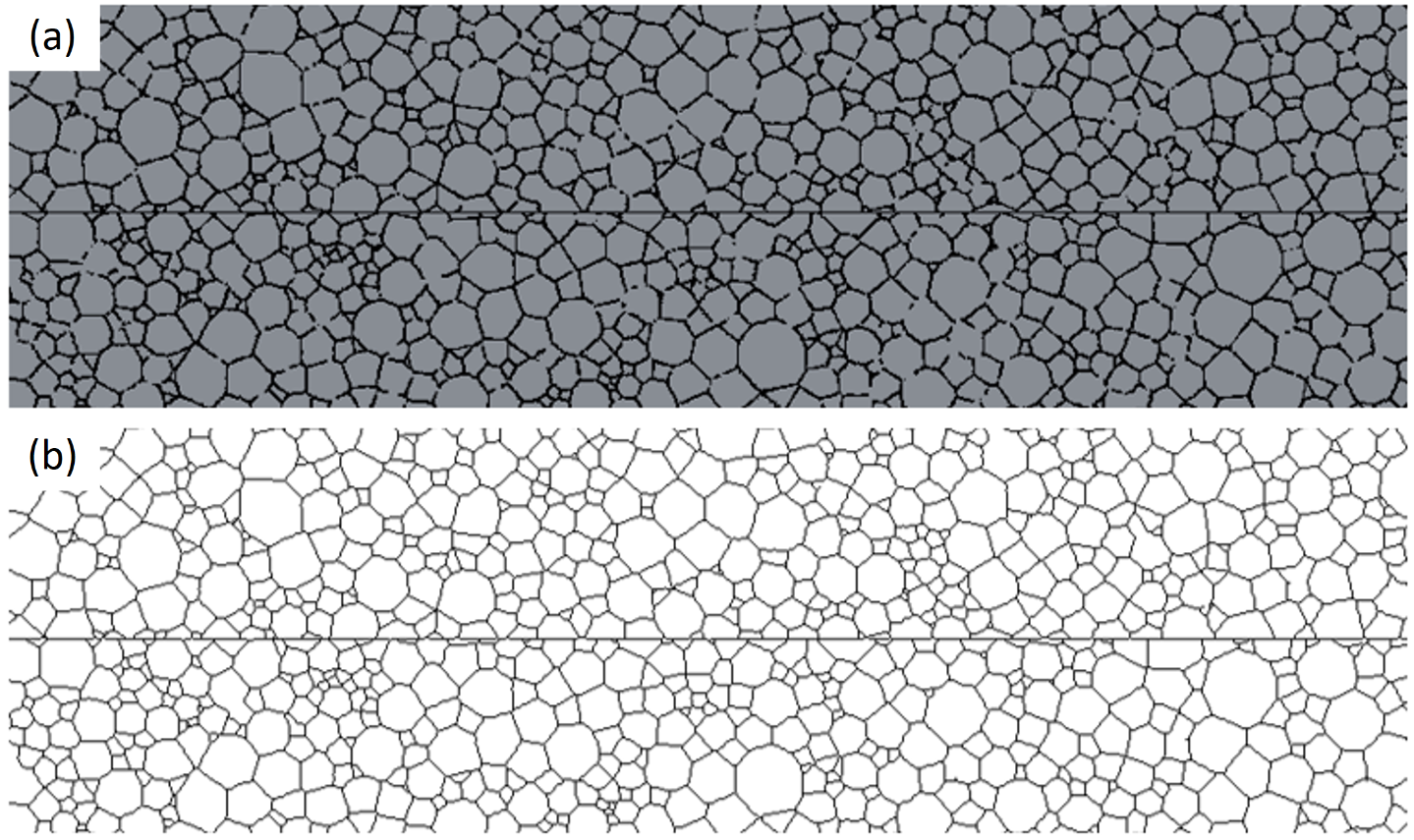}
    \caption{Portion of the image obtained from full-field simulations (a) and portion of the treated image (b)}
    \label{fig:TraitementImage}
\end{figure}

\begin{figure}
 \centering
\includegraphics[width=0.8\textwidth]{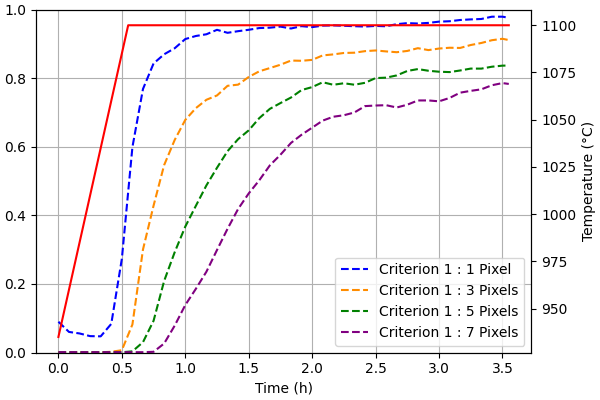}
\caption{$C_1\left(\varepsilon\right)$ for Case I with varying $\varepsilon$ on 3900 initial grains RVE. The red lines correspond to the simulated annealing with the corresponding temperature scale shown on the right side of the chart. }
\label{fig:sensitivity}       
\end{figure}

Once the microstructure generated, the simulations are realized following the level-set numerical framework described previously thanks to the DIGIMU software \cite{Micheli2019,Micheli2023}. Annealing described by the red line in Figure \ref{fig:sensitivity} is considered for all simulations. Every 300s of this diffusion bonding heat treatment, a picture of the microstructure is saved in order to monitor the microstructure evolution at the interface. Image processing is achieved using Fiji and Python that calculate the $C_1\left(\varepsilon\right)$ and $C_2$ criterions for each picture. The image processing steps consist in converting the image to an 8-bit format, cropping it to remove the black background, thresholding to show grain boundaries in black and grains in white, using the watershed and skeletonize functions to close the contours of open grains and to reduce the thickness of grains boundaries. 
An example before and after this processing is presented in Fig.\ref{fig:TraitementImage}.

\begin{figure}
  \centering
    \includegraphics[width=0.8\textwidth]{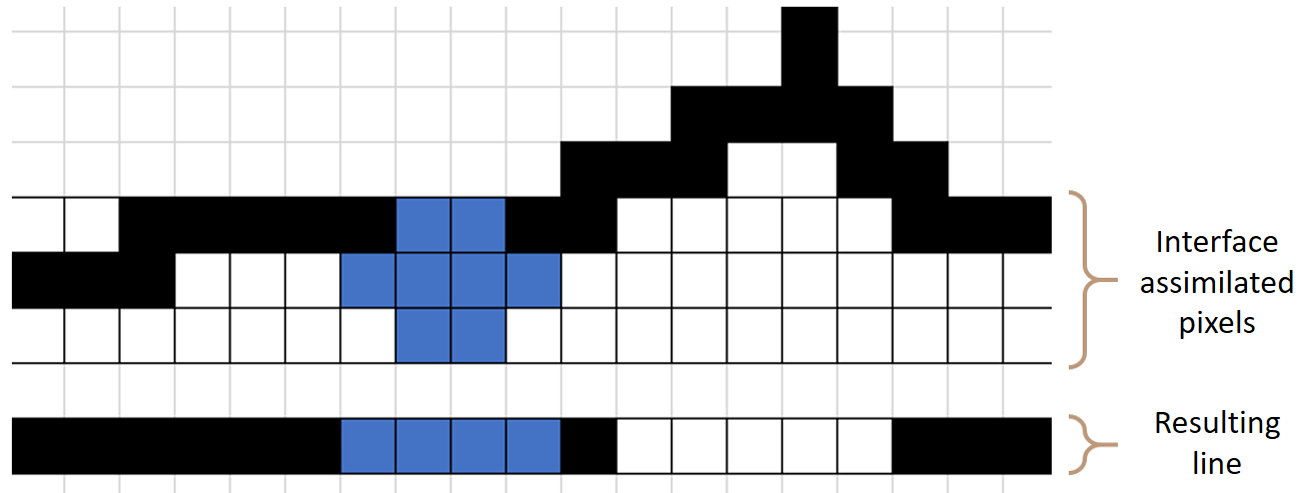}
    \caption{(top) Interface representation, blue pixels represent SPP, black ones represent grain boundary and white ones are bulk of the grains and (bottom) resulting interpretation following the proposed priority rules.}\label{fig:LignePixel}
\end{figure}

To evaluate both interface crossing-criteria on the different cases, an automatic procedure is set up. The interface line can be evaluated on different thicknesses. Fig.\ref{fig:LignePixel} illustrates  an evaluation on a thickness of 3 pixels where black pixels correspond to grain boundaries and blue ones represent second phase particles. For the sake of analyses, a single line is consolidated from the 3 lines evaluated. The pixel state is determined by the state of the 3 corresponding vertical pixels based on order of priority. Second phase particles have the greatest weight, followed by the grain boundaries and then bulk of the grains. The bottom part of the Fig.\ref{fig:LignePixel} illustrates these priority rules. 

A first set of crossing simulations (Case I) has been performed without obstacles to test the crossing criteria behavior : a rectangular RVE of 3900 grains was used to evaluate the first criterion for different numbers of pixels. These preliminary results are displayed in Fig.\ref{fig:sensitivity}.

When evaluated on only one pixel, at the first steps of the simulation, $C_1\left(\varepsilon\right)$ is not null. However, there is not crossing yet, which illustrates that the interface evaluated is to thin. It is clearly apparent from these results that the choice of thickness in terms of pixel count has a significant impact on the resulting curve. Aside from the one-pixel choice, there does not seem to be any clearly superior option, although logically, this analysis thickness should remain small compared to the average grain size. This mainly illustrates the impact of the $\varepsilon$ value which must be fixed to compare the different simulations between them. In the following, this parameter was fixed to $\varepsilon=3$ pixels.

\begin{figure}
\centering
  \includegraphics[width=0.9\textwidth]{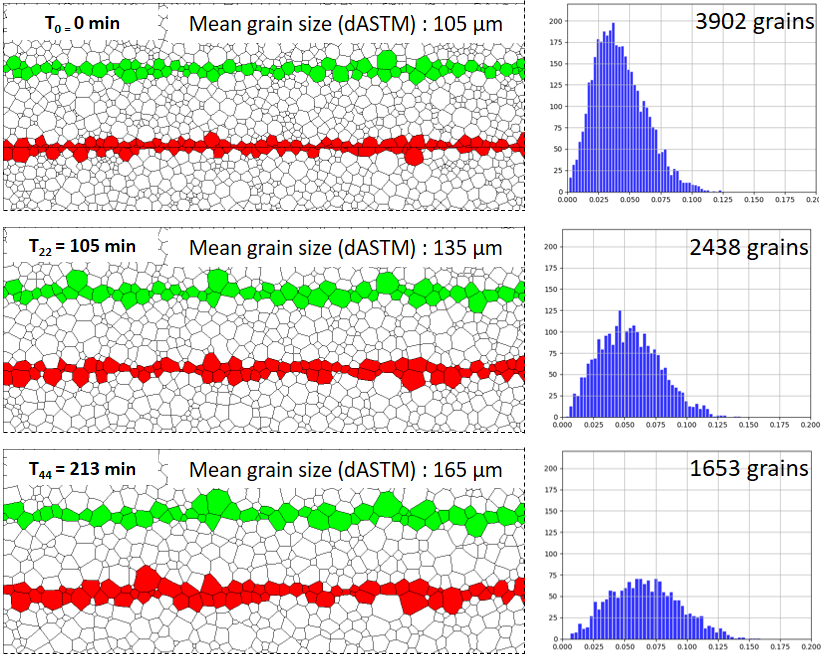}\\
\caption{Representative sections of the RVE at various stages of the thermal cycle, accompanied by the equivalent circle radius (ECR) distribution in mm for each stage. }
\label{fig:CasI}       
\end{figure}

 On Fig.\ref{fig:CasI}, the same sections of the RVE is represented at different stages of the simulated thermal cycle along with the associated equivalent circle radius (ECR) distribution. The chosen stages are the microstructure initial state, an intermediate state after 105 minutes and the final state after 213 minutes. The red colored grains are the ones being in contact or crossing the straight line that was initially the bonding interface. The green colored grains cross an arbitrary chosen line, called the reference line. During the cycle, the mean grain size increases and the histogram spreads. These are phenomena characteristic of grain growth. The interface line, perfectly defined at the initial state, becomes increasingly difficult to distinguish during annealing.

\begin{figure}
\centering
  \includegraphics[width=0.8\textwidth]{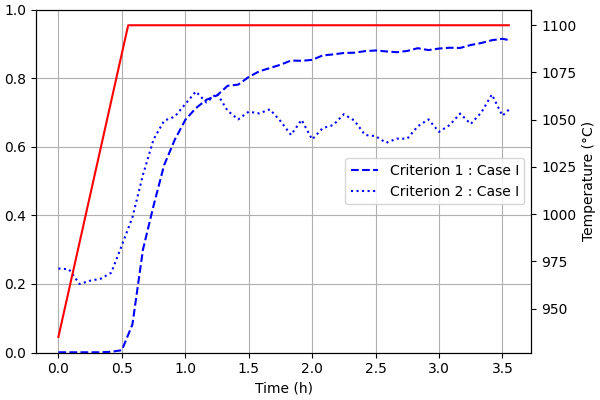}\\
\caption{Crossing evolution evaluated with the criteria $C_1\left(\varepsilon\right)$ and $C_2$ during the thermal cycle for the case I.}
\label{fig:CasIcritere12}       
\end{figure}

The $C_2$ criterion, with the $C_1\left(\varepsilon=3\right)$ criterion already described for this simulation (see Fig.\ref{fig:sensitivity}), are gathered on Fig.\ref{fig:CasIcritere12}. In the first steps, the $C_2$ criterion starts by reducing lightly. It comes from the meshing accuracy. The very fine grains, that are more likely to be present at the interface are not as well defined in the first steps because the meshing is coarser to limit the calculation time during the RVE generation. After several remeshing steps at the beginning of the simulation, this bias tends to disappear. A light improvement of the crossing begins to appear before the temperature reaches the bearing threshold. T-triple junctions on the plane quickly move out of the welding plane to balance themselves according to the Herring equilibrium equation \cite{Herring1999}. As the interface energy $\gamma$ is homogeneous by base hypothesis, the equation simply consists in the angles equality at the multiple junctions. Triple points are thus dragged out of the welding plane and grain growth completes the crossing by homogenisation of the grain size at the interface. 

In the middle of the simulation, it is still possible to distinguish the interface on Fig.\ref{fig:CasI}. Then it becomes increasingly hard because of the grain growth. However, although it is hard to distinguish the interface with the naked eye, the $C_1\left(\varepsilon=3\right)$ criterion curve hardly reaches 1. The $C_2$ criterion indicates that the number of grains crossing or touching the interface remains still 1.5 times greater than the reference one. It seems that the old interface still has an impact on this zone, but that isn’t detected by the $C_1\left(\varepsilon=3\right)$ criterion. The $C_2$ criterion presents many variations between two steps which can be explained by the decrease of thez total number of grain during grain growth. As a perspective of this work, the information quality given by this parameter could be improved by massive simulations with a longer interface portion. 

\begin{figure}
\centering
\includegraphics[width=0.9\textwidth]{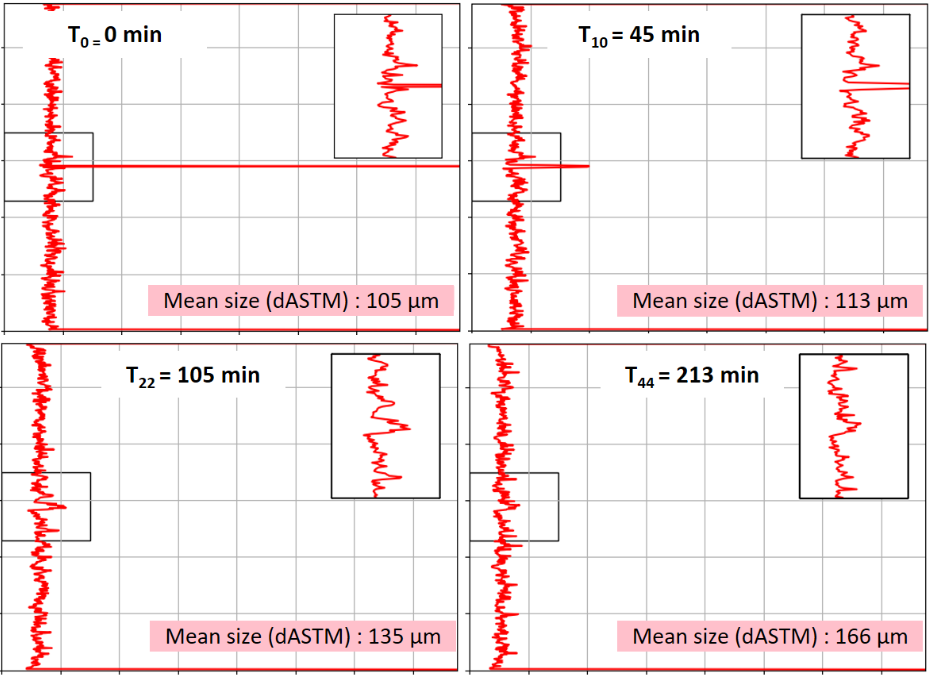}\\
\caption{Evolution of the number of black pixels (grain boundaries) on each image line during various stages of the thermal cycle for Case I.}
\label{fig:EtalementInterface}       
\end{figure}

To be able to estimate the crossing impact on a wider zone, another method is proposed. The number of black pixels on each image line is counted and represented on Fig.\ref{fig:EtalementInterface}. At the initial state, the interface still perfectly defined is represented by a Dirac pic. Gradually the pic representing the interface spreads lightly but the interface pic is then quickly lost in the signal noise. This method is another way to visualize the interface erasing but doesn’t allow precise evaluation of the width affected by the gradual interface erasing. 
The same exploitation method will be applied for the others cases. 

\section{Discussion through a simulation campaign}
\label{ssec:2.2}

Different test cases with second particle populations having different sizes, densities and shapes were realized and are summarized in Tab.\ref{tab:ResuC1}. The Case I corresponds to the test case studied in the previous section. For each case, the size of the RVE, the initial grain size and the characteristics of the second phase particle populations are specified. As the second phase particles are aligned along the welding interface, their density is expressed as a ratio of the total interface length ($f_{SPP}$). Some of them are elongated with an ellipsoidal shape and with their major semi-axis (MSA) parallel to the welding plane. In theses cases, a representative ratio from experimental data of 3 was considered between the MSA and the minor semi-axis. The final $C_1\left(\varepsilon=3\right)$ value is also specified for all configurations. For each case, the horizontal welding interface is still at the center of the RVE while the reference plan (for crossing criteria calculation) is at one quarter of the total height to minimize both border effects and interface proximity. 

\begin{table}
\centering

\caption{List of simulation cases and final $C_1\left(\varepsilon=3\right)$ criterion values.}
\label{tab:ResuC1}       
\begin{tabular}{c|c|c|c|c|c|c}
\hline\noalign{\smallskip}
Case & RVE dim. & dASTM 	& $f_{SPP}$ 		 & SPP 		& shape  	 & $C_1\left(\varepsilon=3\right)$   \\
	 & (mm $\times$ mm)		&   (µm)		&  (\%)				 & MSA (µm)		& ratio &	final value	\\	
\noalign{\smallskip}\hline\noalign{\smallskip}
I 	& 9 x 3 	& 105	&  -			& -			& -		& 0.91 \\
II & 2.1 x 0.8 & 24	&  -			& - 		& -		& 1.00 \\
III  & 1.8 x 0.6 & 23	& 2.6			& 2.4		& 1	& 0.92 \\
IV	& 1.8 x 0.6 & 23	& 6.5			& 2.4		& 1	& 0.74 \\
V  & 1.8 x 0.6 & 24	& 12.5			& 2.3		& 1	& 0.48 \\
VI & 1.8 x 0.6 & 23	& 6.9			& 7.8 	    & 3	& 0.87 \\
VII& 1.8 x 0.6 & 23	& 6.9 (evol.)	& 7.8 	  & 3	& 0.95 \\
\noalign{\smallskip}\hline
\end{tabular}
\end{table}

\subsection{Influence of the initial grain size}

\begin{figure}
\centering

  \includegraphics[width=0.9\textwidth]{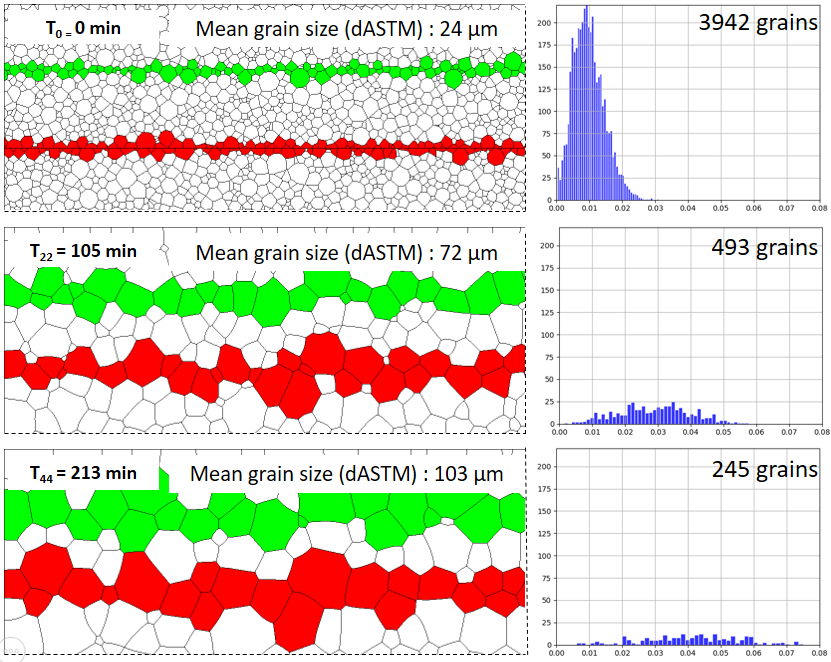}\\

\caption{Partial views of the RVE at various stages of the thermal cycle, accompanied by ECR distributions in mm at different times for the Case II.}
\label{fig:CasII}    
\end{figure}

To study the influence of a thinner initial grain size, Case II with a mean initial grain size (dASTM) of 24 µm can be compared to the Case I. The microstructure pictures on Fig.\ref{fig:CasII}, taken at the same time steps as the previous case (Fig.\ref{fig:CasI}), show that the crossing kinetic is far superior. 

\begin{figure}
\centering
  \includegraphics[width=0.8\textwidth]{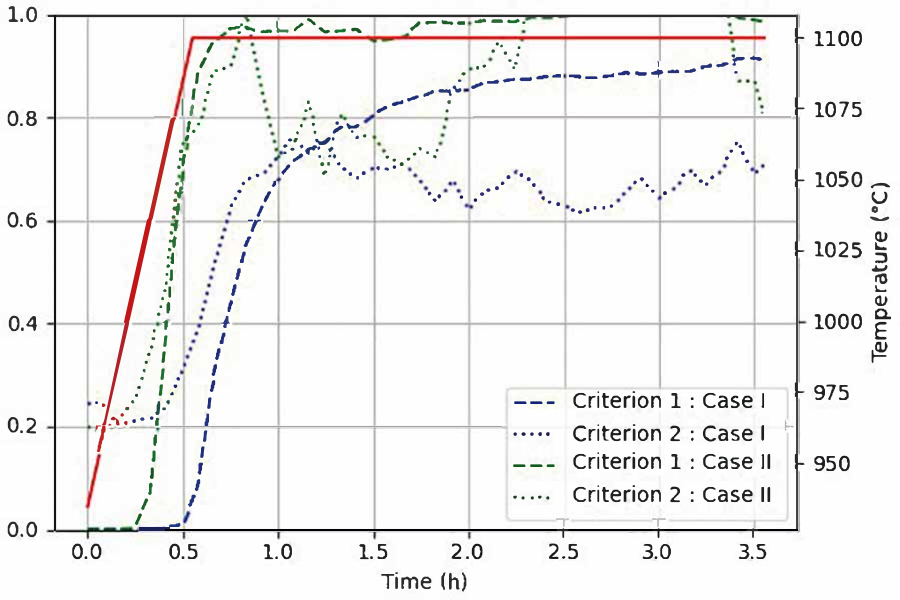}\\
\caption{Crossing evolution evaluated with the criteria $C_1\left(\varepsilon=3\right)$ and $C_2$ during the thermal cycle for the cases I and II.}
\label{fig:CritCasII}       
\end{figure}

The comparison of the crossing evolution, presented on Fig.\ref{fig:CritCasII}, illustrates that the crossing begins before reaching the bearing temperature. The capillary pressure is much greater because the curvature radius of grain boundaries is lesser. As soon as the temperature bearing begins, the $C_1\left(\varepsilon=3\right)$ for the small grains case (Case II) is higher than the maximum value reached with the $C_2$ criterion on the total cycle. The grain number obtained at the end of the simulation is quite low, so the $C_2$ criterion becomes difficult to discuss after 2h of annealing treatment.

In the absence of second phase particles, it seems that if the initial grain size is too big, it is possible that the interface never totally vanishes during the considered long annealing. The difference persists throughout the entire thermal cycle. A better crossing can be obtained by working with finer microstructures. In order to verify whether this capillary pressure increase allows grain boundaries to cross the second phase particles present at the interface, new simulations in presence of obstacles are considered in the following.

\subsection{Influence of the second phase particles density}
\begin{figure}
\centering
  \includegraphics[width=0.9\textwidth]{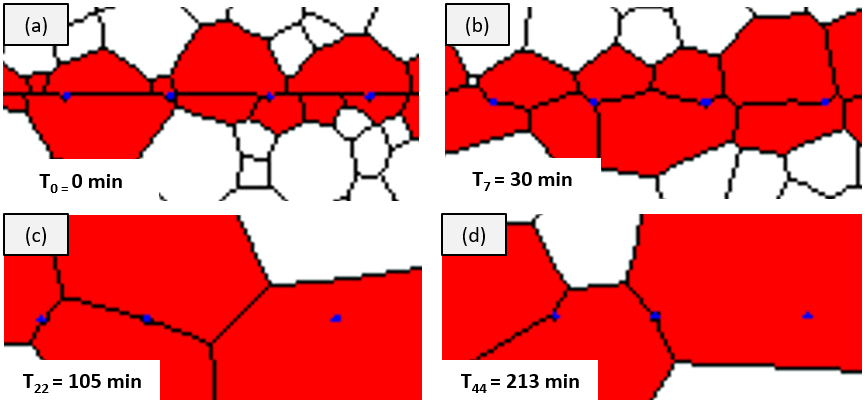}\\
\caption{Partial views of the RVE containing second phase particles at different stages of the thermal cycle for case IV.}
\label{fig:AncrageCasIV}     
\end{figure}

The cases III, IV and V contain circular particles with a diameter from 2.3 to 2.4 µm, that represent respectively, 2.6, 6.5 and 12.5\% of the interface line. The more obstacles there are at the interface, the more they pin the grain boundaries. This effect is illustrated on Fig.\ref{fig:AncrageCasIV} at the state obtained after 30 minutes of thermal treatment. It approximately matches the end of the temperature rise. Some of the particles are crossed, they are then in intragranular position. 

\begin{figure}
\centering
  \includegraphics[width=0.8\textwidth]{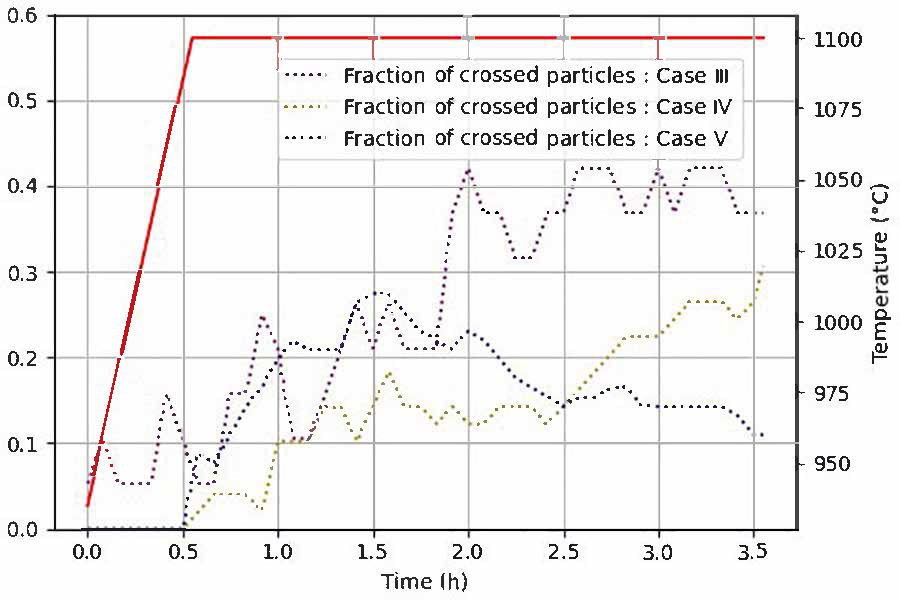}\\
\caption{Number fraction of crossed particles for the cases with circular second phase particles.}
\label{fig:PartIntra}      
\end{figure}

Particle pinning is also a criterion that can provide information on the crossing. The proportion of particles that have been crossed is presented on Fig.\ref{fig:PartIntra}. For case III, which has the lowest particle fraction at the interface, particles are crossed early in the first steps and the amount seems to increase gradually. This trend seems to stagnate after 2 hours of cycle. The variations seem quite sharp but this is due to the fact that the total number of particles is quite low. In the other two cases (Cases IV and V), particles begin to be crossed when the temperature threshold is reached. In case V, where the linear fraction of second phase particles is the highest, particle re-pinning is observed from an hour and a half of the cycle. 

\begin{figure}
\centering
  \includegraphics[width=0.8\textwidth]{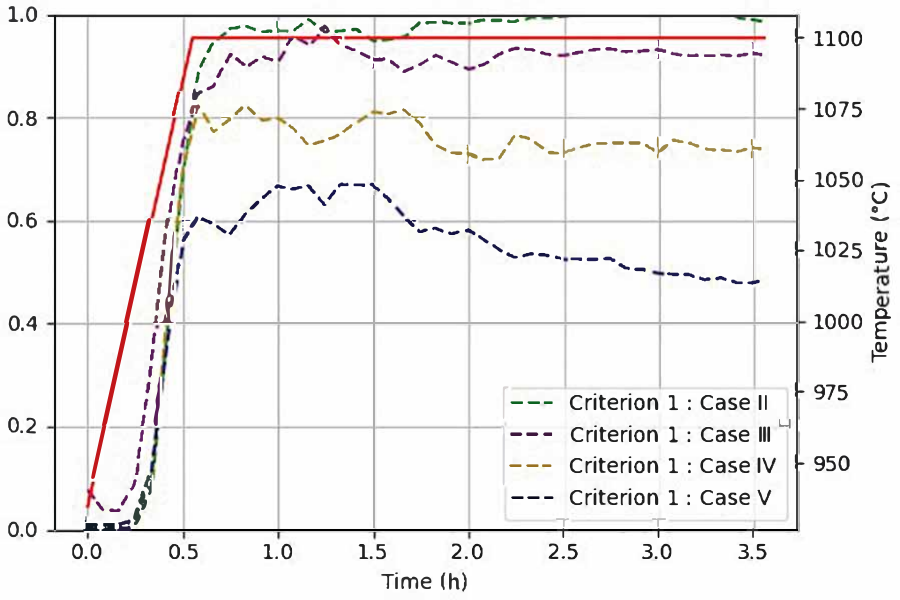}\\
\caption{Criterion $C_1\left(\varepsilon=3\right)$ for the cases with circular second phase particles (Cases III, IV, and V) comparatively to the case with the same mean grain size but without SPP (Case II).}
\label{fig:CompCasII_III_IV_V}      
\end{figure}

There may exist a threshold ratio between grain size and the number of second-phase particles beyond which grain boundary re-pinning becomes observable. In the absence of a triple junction exerting a pulling force on the grain boundary between two adjacent second-phase particles, the most energetically favorable configuration is the straight segment connecting them. As a result, the migration of this portion of the grain boundary is energetically unfavorable, effectively preventing it from being crossed. During grain growth, this can lead to the re-formation of the boundary along the particle interface. This behavior appears to occur in the two cases with the highest obstacle densities (Cases IV and V), as illustrated in Fig.~\ref{fig:CompCasII_III_IV_V}. Notably, the $C_1\left(\varepsilon=3\right)$ value decreases throughout the cycle after an initial rapid increase.

\subsection{Influence of geometry}

\begin{figure}
\centering
  \includegraphics[width=0.8\textwidth]{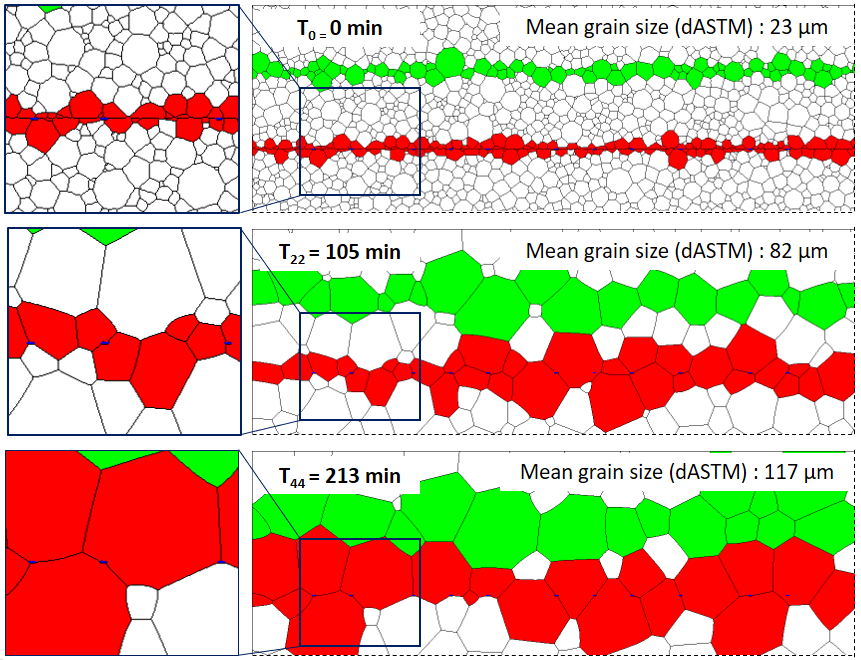}\\
\caption{Partial views of the RVE containing elongated second phase particles at different stages of the thermal cycle for case VI.}
\label{fig:CasVI}     
\end{figure}

Previous cases have been studied with circular particles. However, pore and carbide shapes observed along the bonding interfaces can be much more complex. They seem to extend along the length of the interface. Elliptical particles were therefore also considered at the interface with a shape ratio between their major and minor semi-axis equals to 3. This ratio being considered representative of the material and the conditions studied. The curvature radius is therefore smaller at the ends of the second phase particle. One could expect grain boundaries to remain pinning preferably at this location on the various particles. This can be verified on Fig.\ref{fig:CasVI}.

\begin{figure}
\centering
  \includegraphics[width=0.8\textwidth]{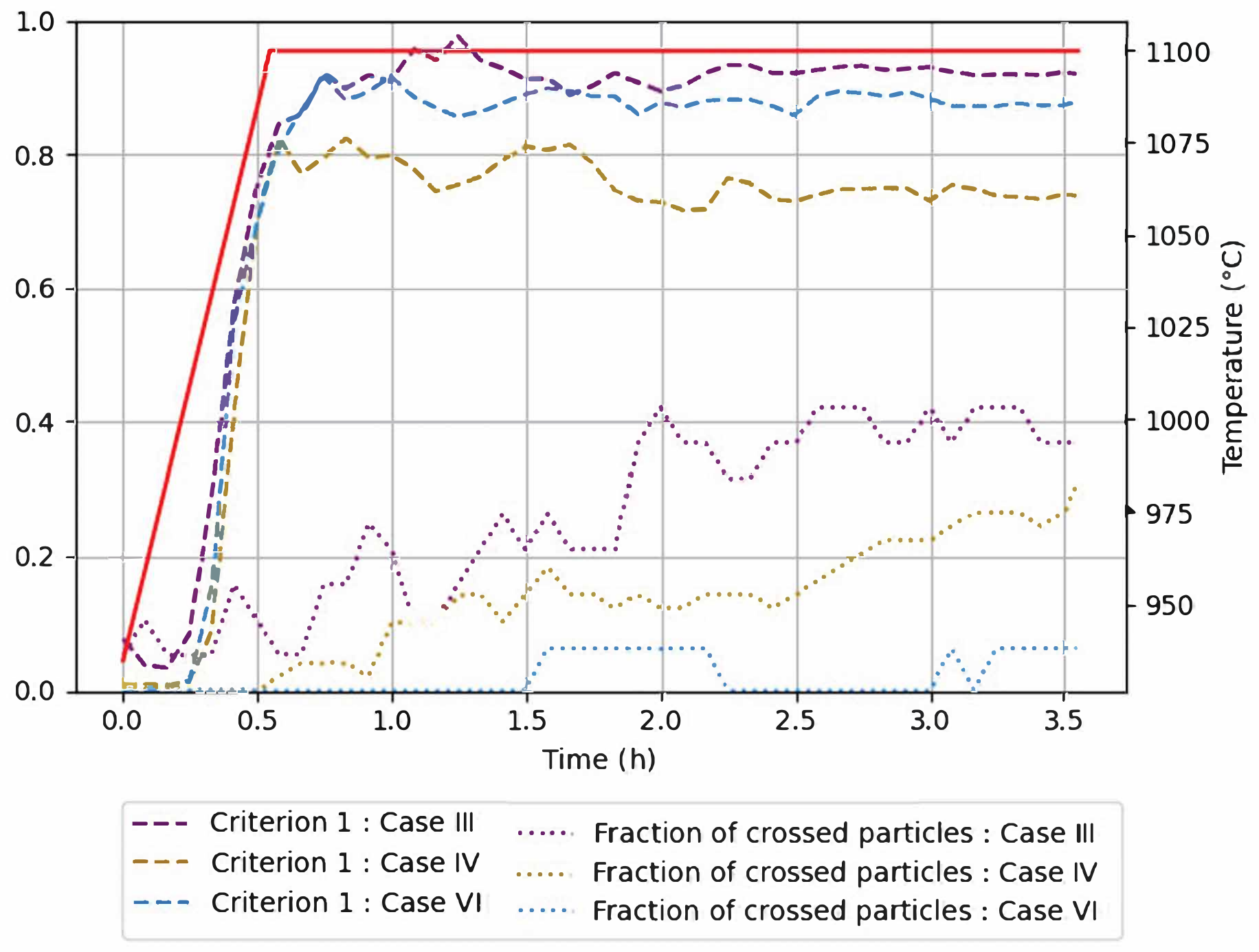}\\
\caption{$C_1\left(\varepsilon\right)$ evolution for the cases containing circular and elliptical particles (cases III, IV, and VI).}
\label{fig:CompCas_III_IV_VI}     
\end{figure}

To assess the influence of particle geometry, this case VI is compared to case IV, which contains a similar linear fraction of defects, and case III, which has a comparable number of second-phase particles at the interface. The results are presented on Fig.\ref{fig:CompCas_III_IV_VI}. At a close linear fraction of obstacles, the case with elliptical particles is the most favorable. However, at a comparable number of obstacles, the case with elliptical obstacles yields poorer results. It appears that the number of pinning points has a more detrimental influence on the quality of the crossing than the linear fraction occupied by the obstacles. However, it still has an impact since the crossing criterion reaches a value closer to 1 for Case IV. It seems that elliptical obstacles parallel to the bonding interface are more negative for the interface passing, meaning that these particles are less likely to be crossed by grain boundaries. The percentage of particles crossed given for each case seems to confirm this hypothesis.

\subsection{Dynamic evolution of obstacles}
\begin{figure}
\centering
  \includegraphics[width=0.8\textwidth]{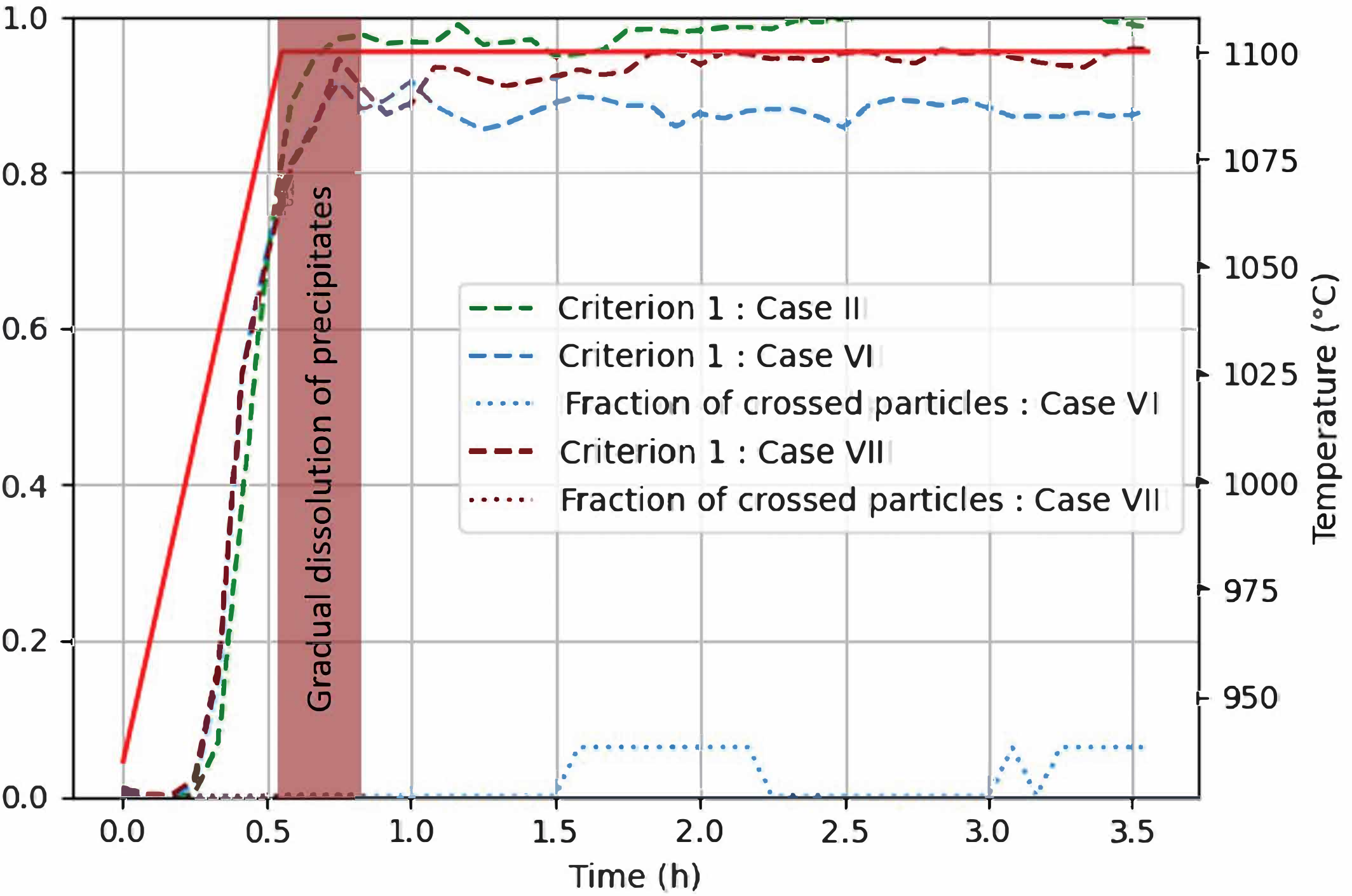}\\
\caption{$C_1\left(\varepsilon=3\right)$ evolution for the cases containing static and dynamic elliptical particles (cases II, VI, VII).}
\label{fig:CompCasII_VI_VII}       
\end{figure}

Finally, during the thermal cycle, the nature and effectiveness of obstacles to boundary migration can evolve. For instance, pores may gradually disappear, and depending on the temperature, new precipitates can form, grow, shrink, or even fully dissolve. To simulate these phenomena, an evolution law can be applied to the second-phase particle population. In the simulated case (Case VII), the second-phase particles were set to dissolve completely at the beginning of the temperature plateau. A realistic kinetics model—undisclosed here for confidentiality reasons—was used to govern pore disappearance based on the material under study. For this case, the same RVE and initial obstacle distribution as in Case VII were used, but with dynamic obstacle behavior. The results are shown in Fig.~\ref{fig:CompCasII_VI_VII}.

It appears that the presence of obstacles that disappear early in the cycle induces a difference in crossing that persists until the end of the three hours of the cycle. Although the dissolution of these obstacles allows an improvement, it does not allow to reach the crossing obtained in the absence of obstacles.

\section{Results Analysis}
\label{ssec:3.1}
Simulations of boundary crossing on perfect interfaces provide an initial critical assessment of the crossing criteria.
The state-of-the-art criterion, $C\left(\varepsilon\right)$, depends on grain size, prompting the exploration of alternative formulations. The first alternative, $C_1\left(\varepsilon\right)$, appears to be more accurate and easier to apply. It corresponds to the ratio of crossings measured along the interface line to those measured along a reference line. Another criterion, $C_2$, was also considered. It is based on the number of grains in contact with the interface relative to those intersecting a reference line. 
While the $C_2$ criterion is appealing because it does not depend on the arbitrary value of $\varepsilon$, which is scale-dependent, it was ultimately not adopted due to its high sensitivity to noise, particularly when the number of simulated grains was low.
The crossing kinetics based on the the $C_1\left(\varepsilon\right)$ criterion appears to be more stable but also highly sensitive to the thickness parameter, $\varepsilon$, which numerically corresponds to the number of pixels used to evaluate the coincident grain boundary length, as illustrated in Fig.~\ref{fig:sensitivity}. While mesh sensitivity is a purely numerical concern, the choice of the parameter $\varepsilon$ is also critical for experimental measurements. It may be linked to the spatial resolution limit and  the characteristic grain size.

\section{Conclusion and perspectives}

Numerous simulations were performed on various cases, starting from either a fine or coarse initial microstructure, with interfaces containing obstacles whose characteristics—such as shape, density, and dynamic behavior— were varied. Among all configurations, the most favorable crossing was observed for a fine initial microstructure and in the absence of obstacles.
From the results involving second-phase particles, circular obstacles at low densities were found to be the least disruptive. In contrast, elongated obstacles aligned with the interface proved more difficult to bypass. One plausible explanation is that the most favorable configuration for overcoming an obstacle involves two triple junctions on either side of the particle, moving perpendicularly to the interface. In such cases, circular particles are more easily crossed than elongated ones aligned with the interface.
In high obstacle density cases, the crossing initially improves but then gradually deteriorates as grain boundaries re-pin on second-phase particles.
As a perspective for future work, it would be valuable to define a predictive parameter—dependent on grain size and the average spacing between precipitates—to estimate the critical grain size at which crossing behavior begins to degrade.

In the current simulation assumptions, it is considered that the densification and/or dissolution of surface species is already complete when the simulation begins. However, in reality, these three phenomena—densification, dissolution, and interface crossing—do not occur strictly sequentially but rather overlap to some extent. This implies that pores, oxides, or other contaminant species may still be present during the early stages of grain boundary crossing. This highlights the importance of continuing work with evolving obstacles.

Initial tests with evolving second-phase particles show that their dissolution during the diffusion bonding cycle improves crossing behavior. However, this improvement still falls short of the performance observed in the absence of precipitates. These simulations help quantify the influence of initial conditions on the final bonding outcome.
One potential goal is to develop a digital twin capable of predicting the final state after a diffusion welding cycle. However, this remains a challenging objective for several reasons. First, the initial simulation state only approximates reality, and in practice, densification, dissolution, and interface crossing occur simultaneously.
Moreover, there are limitations related to computational cost. To keep simulations tractable, the ratio between precipitate size and the simulated domain must not exceed four orders of magnitude in typical level-set simulations at the polycrystalline scale. Accurate results also require a large number of grains to be preserved during the simulation. To properly capture small obstacles, the mesh must be fine enough to resolve their interfaces, which is even more demanding when obstacles evolve over time.
Finally, extending the simulations to 3D significantly increases computational requirements. One promising avenue, to overcome these current limits, is the recently developed front-tracking ToRealMotion approach \cite{Florez2020}, which allows for the precise modeling of complex second-phase particle populations—even with a large number of grains—while maintaining the accuracy of the level-set method \cite{Florez2025}.

\section*{Acknowledgements}
The authors thank the CEA-Liten and the CALHIPSO Equipex+ project (grant ANR-21-ESRE-0039) for providing support and hosting this study. The authors also thanks the DIGIMU consortium partners and the French Research National Agency (ANR), which, through the RealIMotion Chair (grant ANR-22-
CHIN-0003), have contributed to the development of the numerical framework used in this study.

\bibliography{biblio}

\end{document}